\documentclass[11pt]{article}
\pdfoutput=1
\usepackage{dcolumn}% Align table columns on decimal point
\usepackage{bm}% bold math
\usepackage{graphicx}% Include figure files
\usepackage{amssymb,amsmath}
\usepackage{multirow}
\usepackage{cite,color,url}
\usepackage[colorlinks=true
,urlcolor=blue
,anchorcolor=blue
,citecolor=blue
,filecolor=blue
,linkcolor=blue
,menucolor=blue
,linktocpage=true
,pdfproducer=medialab
,pdfa=true
]{hyperref}

\usepackage{slashed}
\usepackage{epsfig,psfrag,rotating,soul}
\usepackage{rotfloat}
\usepackage[font={small}]{caption}
\usepackage{colortbl}
\evensidemargin \oddsidemargin
\allowdisplaybreaks

\let\OLDthebibliography\thebibliography
\renewcommand\thebibliography[1]{
  \OLDthebibliography{#1}
  \setlength{\parskip}{0pt}
  \setlength{\itemsep}{0pt plus 0.3ex}
}

\definecolor{mygray}{gray}{0.85} 
\definecolor{myblue}{cmyk}{0.65, 0.37, 0.0, 0.19}

\begin{document}
\thispagestyle{empty}

\def\thefootnote{\fnsymbol{footnote}}

\begin{flushright}
IFT-UAM/CSIC-21-44
\end{flushright}

\vspace*{1cm}

\begin{center}

\begin{Large}
\textbf{\textsc{Novel Higgsino Dark Matter Signatures at the LHC}}
\end{Large}

\vspace{1cm}

{\sc
Ernesto~Arganda$^{1, 2}$%
\footnote{{\tt \href{mailto:ernesto.arganda@csic.es}{ernesto.arganda@csic.es}}}%
, Antonio~Delgado$^{3}$%
\footnote{{\tt \href{mailto:adelgad2@nd.edu}{adelgad2@nd.edu}}}%
, Roberto~A.~Morales$^{1}$%
\footnote{{\tt \href{mailto:robertoa.morales@uam.es}{robertoa.morales@uam.es}}}%
,		 Mariano~Quir\'os$^{4}$%
\footnote{{\tt \href{quiros@ifae.es}{quiros@ifae.es}}}%

}

\vspace*{.7cm}

{\sl
$^1$Instituto de F\'{\i}sica Te\'orica UAM/CSIC, \\
C/ Nicol\'as Cabrera 13-15, Campus de Cantoblanco, 28049, Madrid, Spain

\vspace*{0.1cm}

$^2$IFLP, CONICET - Dpto. de F\'{\i}sica, Universidad Nacional de La Plata, \\ 
C.C. 67, 1900 La Plata, Argentina

\vspace*{0.1cm}

$^3$Department of Physics, University of Notre Dame, 225 Nieuwland Hall \\
Notre Dame, IN 46556, USA

\vspace*{0.1cm}

$^4$Institut de F\'{\i}sica d'Altes Energies (IFAE) and BIST, Campus UAB \\
08193, Bellaterra, Barcelona, Spain

}

\end{center}

\vspace{0.1cm}

\begin{abstract}
\noindent
In the LHC searches for gluinos it is usually assumed that they decay predominantly into the lightest neutralino plus jets. In this work we perform a proof-of-concept collider analysis of a novel supersymmetric signal in which gluinos decay mostly into jets and the bino-like neutralino ($\tilde\chi_3^0$), which in turn decays into the lightest Higgsino-like neutralino ($\tilde\chi_1^0$), considered the dark matter candidate, together with the SM-like Higgs boson ($h$). This new physics signal then consists of an LHC final state made up by four light jets, four $b$-jets, and a large amount of missing transverse energy. We identify $t \bar t$, $V$+jets ($V$= $W$, $Z$), and $t \bar t + X$ ($X$ = $W$, $Z$, $\gamma^*$, $h$) productions as the most problematic backgrounds, and develop a search strategy for the high luminosity phase of the LHC, reaching signal significances at the evidence level for a luminosity of 1000 fb$^{-1}$. The prospects for a luminosity of 3000 fb$^{-1}$ are even more promising, with discovery-level significances.
\end{abstract}

\def\thefootnote{\arabic{footnote}}
\setcounter{page}{0}
\setcounter{footnote}{0}

\newpage

\section{Introduction}
\label{intro}

After the Higgs boson discovery~\cite{Aad:2012tfa,Chatrchyan:2012ufa} at the LHC lots of efforts of the CMS and ATLAS collaborations are in searches for physics beyond the Standard Model (BSM). So far the results have been null so bounds are put in popular models albeit there are caveats on those bounds. The reinterpretation of the searches are normally done in the context of simplified models where it is easier to draw conclusions. An example of those situations are gluino searches done at the LHC (for a recent summary, see, for instance,~\cite{MoriondEW2021} and~\cite{MoriondQCD2021}). In most of the cases it is assumed that the gluino decays with a branching ratio equal to 1 to the lightest neutralino plus jets, which in fact makes an implicit assumption on the supersymmetric (SUSY) spectrum and couplings. If this assumption is not fulfilled many experimental bounds could be evaded. It is thus interesting to explore other (less conventional) possibilities, as very often they are theoretically well motivated, as it is the case we will explore in this paper.

In this work we develop a search strategy for a novel signature at the LHC of Higgsino dark matter, proposed in~\cite{Delgado:2020url}, where the gluino will not decay predominately to the lightest neutralino plus jets. Under very general conditions, that will be explained in section~\ref{th-frame}, there could be several electroweakinos lighter than the gluino, which will change dramatically the signatures at the LHC. The aim of our analysis is more to give a proof of principle, than providing an elaborated strategy, to show which kinematical variables and cuts may be effective for this kind of scenarios. 
Let us finally emphasize that, in general, it is very important for the next run of the LHC to go beyond the usual simplified models, and to design searches, to look for kinematic variables and to optimize cuts, to be sensitive to more scenarios than just the ones captured by simplified models or spectra.

The rest of the paper is organized as follows. The general theoretical framework for the model we will consider is provided in Sec.~\ref{th-frame}. In this framework our guideline will be the possibility of having a 1.1 TeV Higgsino as dark matter. The collider analysis will be done in Sec.~\ref{collider} while our conclusions will be drawn in Sec.~\ref{conclus}.

\section{Theoretical Framework}
\label{th-frame}

Identifying the lightest neutralino $\tilde\chi_1^0$ as the lightest supersymmetric particle (LSP), and thus a dark matter candidate in the presence of $R$ parity~\cite{Jungman:1995df}, is one of the most appealing features of the minimal supersymmetric extension of the Standard Model (MSSM)~\cite{Nilles:1983ge,Haber:1984rc,Gunion:1984yn}. Given the strong LHC bounds on the mass of supersymmetric particles, and the plethora of null results on dark matter direct searches, there remains a preferred supersymmetric scenario: an almost pure Higgsino with a mass $\sim 1.1$ TeV~\cite{Roszkowski:2017nbc,Kowalska:2018toh}. This requirement (almost) fixes the theoretical framework in the electroweakino (neutralino/chargino) sector as it generically requires that $\mu\sim 1.1$ TeV (where $\mu$ is the supersymmetric Higgsino mass) while $M_1,\, M_2 \gg \mu$ (where $M_1$ and $M_2$ are soft supersymmetry breaking Majorana masses for the fermionic partners of the $U(1)_Y$ and $SU(2)$ gauge bosons, bino, and wino respectively). 

The Majorana masses $M_{1,2}$ are defined at the low scale and their values depend on the mechanism of supersymmetry breaking. While the requirement of the Higgsino being the LSP rules out gauge mediation (for which the gravitino is the LSP) as the transmission mechanism for supersymmetry breaking, gravity mediation seems to be the preferred one, as there is room for the lightest neutralino to be the LSP and moreover the supersymmetric mass $\mu$ can be generated by the Giudice-Masiero mechanism~\cite{Giudice:1988yz}. In gravity mediation, all supersymmetry breaking parameters, and in particular $M_{1,2}$ are generated at the high (unification) scale, i.e.~$M_{1,2}^0$, and their value at low scale is obtained by means of the renormalization group equation (RGE) running. Unification conditions are usually assumed, i.e.~$M_1^0=M_2^0$, but even assuming that $M_1^0\sim M_2^0$, after the RGE running we have $M_2\sim 2 M_1$ so that the bino $\tilde\chi_3^0$ is lighter than the wino $\tilde\chi_4^0$. 

Under these circumstances the neutralino sector is almost completely fixed: \textit{i)} There are two (almost) purely Higgsinos, $\tilde\chi_1^0,\, \tilde\chi_2^0$, with masses $\sim 1.1$ TeV and a mass separation of a few GeV. \textit{ii)} There is a bino $\tilde\chi_3^0$ with a mass $m_{\tilde\chi_3^0}\sim M_1$ and a wino with a mass $m_{\tilde\chi_4^0}\sim 2 m_{\tilde\chi_3^0}$. At the same time the constraints from the XENON1T experiment on direct detection~\cite{Aprile:2018dbl}, analyzed in Ref.~\cite{Kowalska:2018toh}, put the constraint, for the case of equal masses at the unification scale,  $M_1^0=M_2^0\gtrsim 3.2$ TeV, which translates into the lower bounds $m_{\tilde\chi_3^0}\gtrsim 1.5$ TeV and $m_{\tilde\chi_4^0}\gtrsim 2.7$ TeV~\cite{Delgado:2020url}. As for the chargino sector, the lightest state $\tilde\chi_1^\pm$ is almost degenerate with the LSP, with a few GeV gap, while the heaviest chargino is almost degenerate with the heavy neutralino, so that $m_{\tilde\chi_2^\pm}\gtrsim 2.7$ TeV.

\begin{figure}[ht!]
	\centering
		\includegraphics{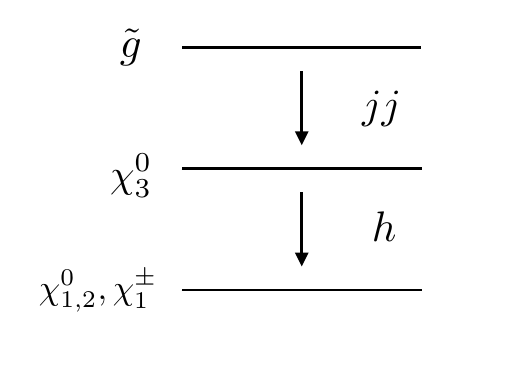} 
			\caption{\it Typical spectrum considered in the analysis with the decay channels shown close to the arrows.}
		\label{fig:spectrum}
\end{figure}

On the other hand the gluino $\tilde g$ mass $M_{\tilde g}$ is also fixed by the breaking mass $M_3^0$ at the unification scale. In our theoretical framework the gluino mass is not unified with the electroweak masses $M_{1,2}^0$ so that it will be considered as a free parameter. This is a safe assumption as the gluino mass does not enter the process of electroweak breaking at the tree level. We will assume that the gluino mass will be close to its present experimental bound $M_{\tilde g}\sim 2$ TeV. Moreover we are going to assume, for simplicity, that all other sparticles including squarks are more massive than the gluino, nonetheless all decays are assumed to be prompt. In this case the possible channels for the gluino decay are $\tilde g\to\tilde\chi_{1,2}^0jj$,  $\tilde g\to \tilde\chi_{1}^\pm jj$, and  $\tilde g\to \tilde\chi_{3}^0jj$, mediated by the decay $\tilde g\to \tilde q^*_a q_a$, where $a$ is a generation label, and followed by $\tilde q^*_a\to \tilde\chi_{1,2}^0 q_a$, $\tilde q^*_a\to \tilde\chi_{1}^\pm q_a$ (induced by the Yukawa coupling $y_{q_a}$) and $\tilde q^*_a\to \tilde\chi_{3}^0 q_a$ (induced by the $U(1)$ gauge coupling $g_1$), respectively. The typical situation that current analyses consider and cover is that the direct decay to nearly degenerate Higgsinos dominates ($\chi_{1,2}^0$,$\chi_1^\pm$). If, instead, the gluino decays predominantly to $\chi^0_3$, one will get a final state with several energetic jets and $b$-quarks that will evade current bounds. The decay channels of the gluino depend on the details of the squark spectrum: if the first two generations of squark are less massive than the third generation, then the decay to $\chi_3^0$ is favored, being of electroweak nature as opposed to the decay to the Higgsino which is proportional to the corresponding Yukawa coupling. In Fig.~\ref{fig:spectrum} we have a schematic view of the spectrum and decays that are going to be analyzed in the next section.

\section{Collider Analysis}
\label{collider}

The experimental signature under study at the LHC comes from the SUSY production of a pair of gluinos, $pp \to \tilde g \tilde g$, that decay into $\tilde \chi_3^0$ and two light jets ($\tilde g \to \tilde \chi_3^0 j j$). We consider then that each $\tilde \chi_3^0$ decays into the LSP ($\tilde \chi_1^0$) and the lightest MSSM Higgs boson, $h$, identified as the 125-GeV SM-like Higgs boson discovered at the LHC, which decays into a pair of $b$-quarks. Therefore, the final state is made of four light jets, four $b$-jets, and a large amount of missing transverse energy ($4 j+ 4 b + E_T^\text{miss}$), whose main SM backgrounds are QCD multijet; $Z$ + jets and $W$ + jets productions; $t \bar t$ production; $t \bar t$ production in association with electroweak or Higgs bosons, $t \bar t$ + $X$ ($X$ = $W$, $Z$, $\gamma^*$, $h$); and diboson production ($WW$, $ZZ$, $WZ$, $Wh$, and $Zh$) plus jets.

We develop our search strategy for a LHC center-of-mass energy of $\sqrt{s}$ = 14 TeV and a total integrated luminosity of $\cal{L}$ = 1000 fb$^{-1}$, compatible with the high-luminosity LHC (HL-LHC) phase. We make use of {\tt MadGraph\_aMC@NLO 2.7}~\cite{Alwall:2014hca} for the Monte Carlo generation of both signal and background events, whose parton shower and hadronization is performed with {\tt PYTHIA 8}~\cite{Sjostrand:2014zea}, while the detector response simulation is achieved with {\tt Delphes 3}~\cite{deFavereau:2013fsa}. From the proposed new physics signal, one would expect in the final state very energetic light jets and $b$-jets, coming from the decays of gluinos and Higgs bosons, respectively. Therefore, with the intention of reducing the large background cross sections and making event generation more efficient, we impose the following generator-level cuts on the $p_T$ of the light jets and $b$-jets for the background simulation~\footnote{For the signal simulation, we use the default cuts on the $p_T$ of the light jets and $b$-jets ($p_T^j >$ 20 GeV and $p_T^b >$ 20 GeV).}:
\begin{eqnarray}
p_T^{j_1} > 180 \, \text{GeV} \,, \quad p_T^{j_2} > 140 \, \text{GeV} \,, \quad p_T^{j_3} > 70 \, \text{GeV} \,, \quad p_T^{j_4} > 35 \, \text{GeV} \,, \nonumber\\
p_T^{b_1} > 90 \, \text{GeV} \,, \quad p_T^{b_2} > 20 \, \text{GeV} \,, \quad p_T^{b_3} > 20 \, \text{GeV} \,, \quad p_T^{b_4} > 20 \, \text{GeV} \,,
\label{generatorcuts}
\end{eqnarray}
where $j_1\dots j_4$ ($b_1 \dots b_4$) runs from the most to the least energetic light ($b$-) jet. 
Dealing with many jets in the final state, the MLM algorithm~\cite{Mangano:2002,Mangano:2006rw} was implemented for jet matching and merging. In order to optimize the simulation and checking that the jet related distributions are smooth, the {\tt xqcut} was set to 20 for all simulated samples and {\tt qcut} equal to 550, 50, and 30 for signal, $t \bar t$-like and backgrounds with bosons, respectively. 

With this in mind, the following comments on the signal and backgrounds are pertinent:
\begin{itemize}
\item The SUSY spectrum and branching ratios for the signal benchmark have been computed with {\tt SOFTSUSY}~\cite{Allanach:2001kg,Allanach:2017hcf,Allanach:2013kza,Allanach:2009bv,Allanach:2011de,Allanach:2014nba,Allanach:2016rxd}, while the production cross section of a pair of gluinos is obtained from~\cite{LHCSUSYxs}. The relevant mass parameters of our benchmark for the proposed SUSY signature are $M_{\tilde g}$ = 2.1 TeV, $m_{\tilde \chi_3^0}$ = 1.6 TeV, and $m_{\tilde \chi_1^0}$ = 1.2 TeV, with the first two generations of squarks at masses around $\sim$ 4 TeV and the third generation of squarks decoupled. The corresponding gluino-pair production cross section and branching ratios are $\sigma(pp \to \tilde g \tilde g)$ = 1.1 fb, BR($\tilde g \to \tilde \chi_3^0 j j$) = 0.82, BR($\tilde \chi_3^0 \to \tilde \chi_1^0 h$) = 0.27, and BR($h \to b \bar b$) = 0.58. With these values, 20 signal events are expected for $\cal{L}$ = 1000 fb$^{-1}$.
\item The QCD multijet background is unmanageable with our computational capacity, and is usually treated with data-driven techniques. In our case, taking into account that our signal will have a large amount of $E_T^\text{miss}$, variables related to this observable, such as the $E_T^\text{miss}$ significance, greatly reduce this class of backgrounds with instrumental missing transverse energy, bringing practically to zero the number of expected events. Therefore, we can consider the QCD multijet background as negligible and it will not be included in our analysis.
\item Regarding the $V$+jets production, including both $Z$+jets and $W$+jets, we considered a pair of $b$-jets and a pair of light jets leading to four extra jets and a genuine source of missing energy through neutrinos coming from the decay of the gauge bosons (with BR($Z\to\nu\nu$) = 0.2 and BR($W\to l\nu)$ = 0.21). Other combinations of extra jets do not have $b$ or light jets enough and more than 4 extra jets are out of our simulation capacity. Then, taking into account the generator setup, we expect $5.6 \times 10^4$ for $Z$+jets and $3 \times 10^5$ events for $W$+jets with $\cal{L}$ = 1000 fb$^{-1}$.
\item Related to the $V$+jets background, the diboson production can be safely neglected in this analysis since it is subdominant with an amount of roughly $10^{-3}$ times the $V$+jets (which we will see it is already under control).
\item The $t\bar{t}$ production, with both fully-hadronic and semileptonic decay channels, is the most dangerous background. The corresponding branching fractions are BR($t\bar{t}_{\rm had}$) = 0.457 and BR($t\bar{t}_{\rm semilep}$) = 0.438. After the generator-level cuts, we expect $1.36 \times 10^6$ and $0.42 \times 10^6$ events, respectively. We also consider one extra jet in the simulation, resulting in $0.83 \times 10^6$ and $0.25 \times 10^6$ events more for the hadronic and semileptonic channels, respectively.
\item Concerning the $t\bar{t} + X$ backgrounds, even though is much smaller than the $t\bar{t}$ ones, the extra boson provide genuine source of missing energy (more $b$-jets) for the hadronic (semileptonic) top-quark pair. Explicitly, we consider $t\bar{t}_{\rm had}+(Z\to\nu\nu)$, $t\bar{t}_{\rm had}+(W\to l\nu)$, $t\bar{t}_{\rm semilep}+(Z\to b\bar{b})$, $t\bar{t}_{\rm semilep}+(\gamma^*\to b\bar{b})$, and $t\bar{t}_{\rm semilep}+(h\to b\bar{b})$. We also include one extra jet to each process, leading to $2.9\times 10^3$ expected events in this category.
\end{itemize}
Next we will perform a characterization of the signal against the dominant SM backgrounds in order to define the most promising signal regions for our search strategy.
In our analysis, the previously defined backgrounds are separated in four categories: $t\bar{t}_{\rm had}+j$ (inclusive), $t\bar{t}_{\rm semilep}+j$ (inclusive), $V$+jets, and $t\bar{t}+X+j$ (inclusive).

\begin{figure}%[ht!]
	\begin{center}
		\begin{tabular}{cc}
			\centering
			\hspace*{-3mm}
			\includegraphics[scale=0.4]{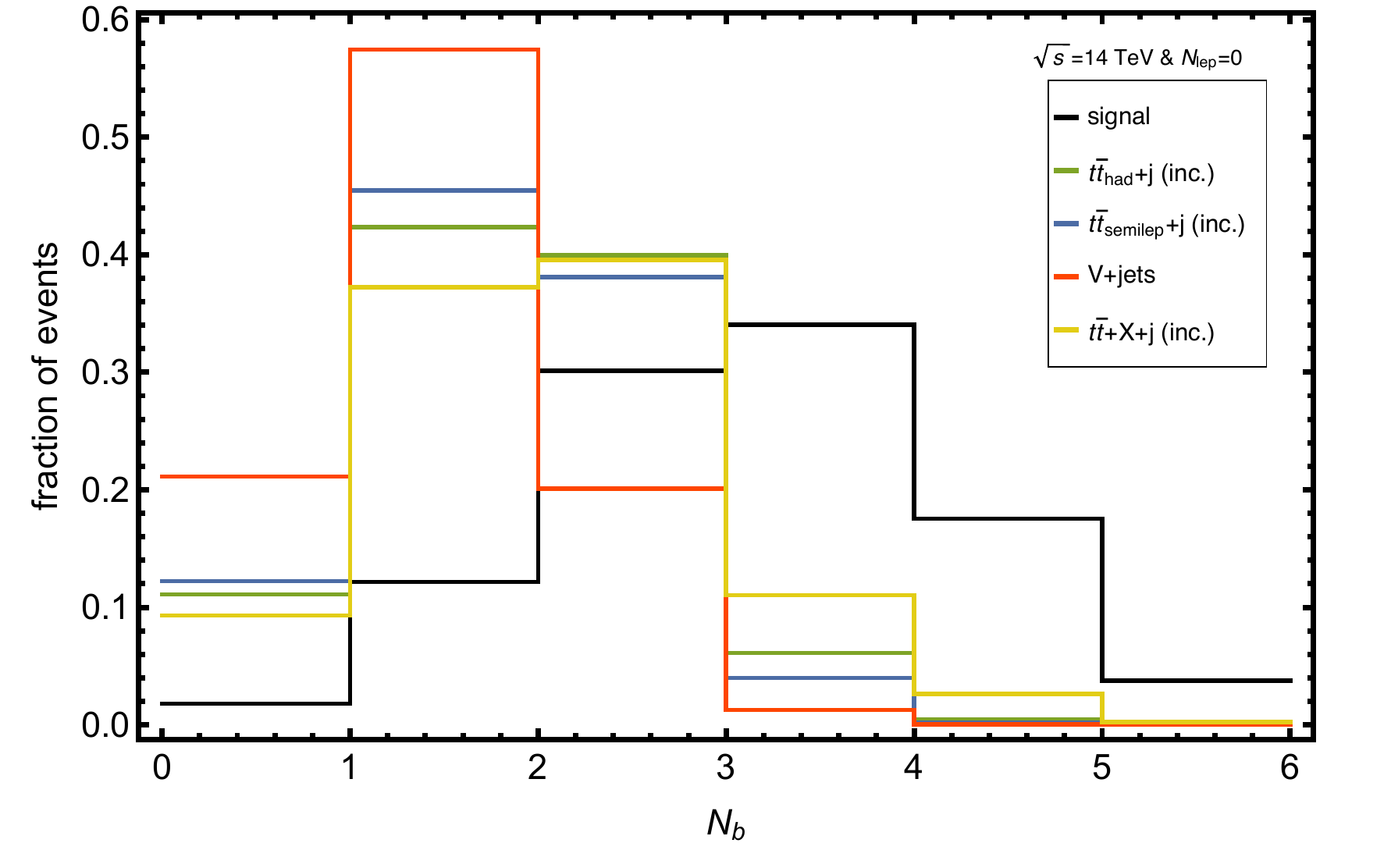} &
			\includegraphics[scale=0.4]{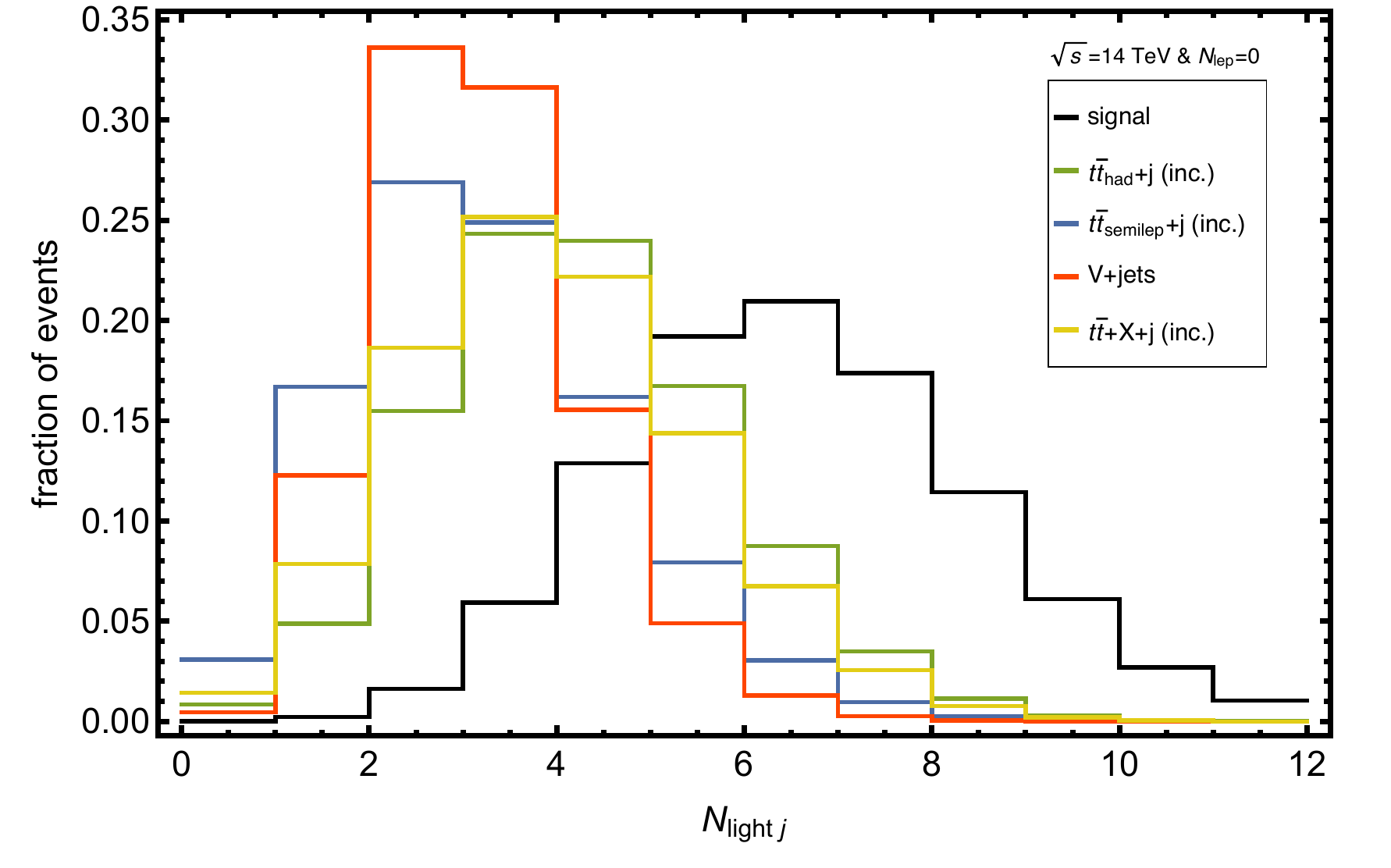}	
		\end{tabular}
		\caption{\it Distributions (with a lepton veto) of the fraction of signal and background events of the number of identified $b$-jets $N_b$ (left panel) and the number of light jets $N_j$ (right panel).}
		\label{fig:Nb-Nj}
	\end{center}
\end{figure}

In Fig.~\ref{fig:Nb-Nj} we depict the distributions of the fraction of signal and backgrounds events of the number of identified $b$-jets $N_b$ (left panel) and the number of light jets $N_j$ (right panel). In order to avoid one of the most dangerous background, the semileptonic $t \bar t$ production, we firstly set a lepton veto ($N_\ell$ = 0), which have been already imposed on the distributions on both plots of Fig.~\ref{fig:Nb-Nj}. One of the most challenging task of the proposed signature is the identification of $b$-jets, since the signal is characterized by 4 bottom quarks coming from the Higgs boson decays. It is clear from the left panel of Fig.~\ref{fig:Nb-Nj} that the requirement of identifying 4 $b$-jets would reduce the number of signal events to less than half. Therefore, we are going to impose two class of selection cuts related to the number of identified $b$-jets: a {\it loose} cut with at least 2 $b$-jets in the final state ($N_b \geq$ 2) and a {\it tight} cut, requiring at least 3 reconstructed $b$-jets ($N_b \geq$ 3). The signal consists also of 4 light jets, then we add to the selection-cut set the requisite of having at least 4 light jets in the final state ($N_j \geq$ 4). Thus, the selection cuts that characterize our signal are as follows:
\begin{eqnarray}
loose: \quad N_b \geq 2 \,, \quad N_j \geq 4 \,, \quad N_\ell = 0 \,, \nonumber\\
tight: \quad N_b \geq 3 \,, \quad N_j \geq 4 \,, \quad N_\ell = 0 \,.
\label{SRdefinitions}
\end{eqnarray}

\begin{figure}%[ht!]
	\begin{center}
		\begin{tabular}{cc}
			\centering
			\hspace*{-3mm}
			\includegraphics[scale=0.4]{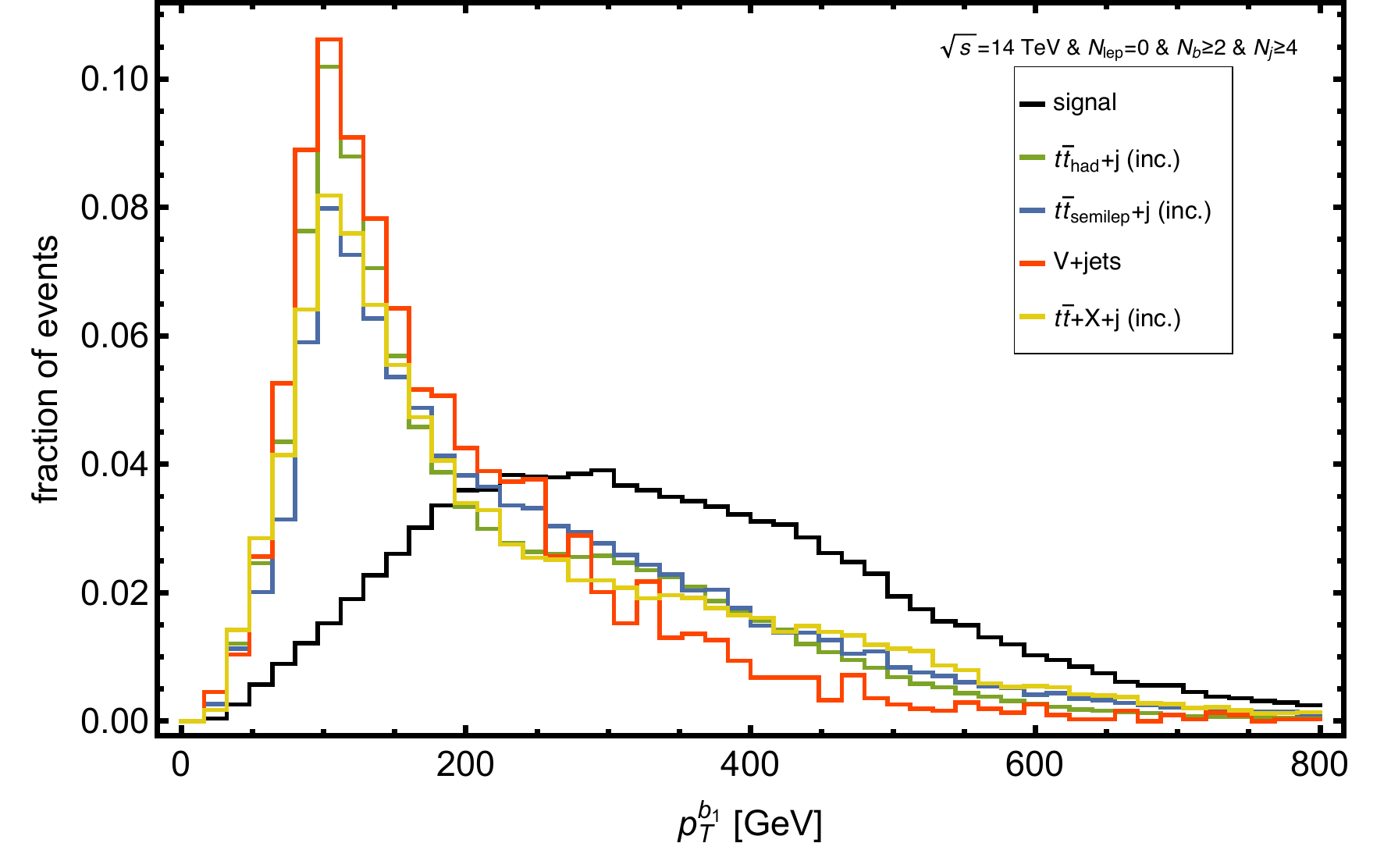} &
			\includegraphics[scale=0.4]{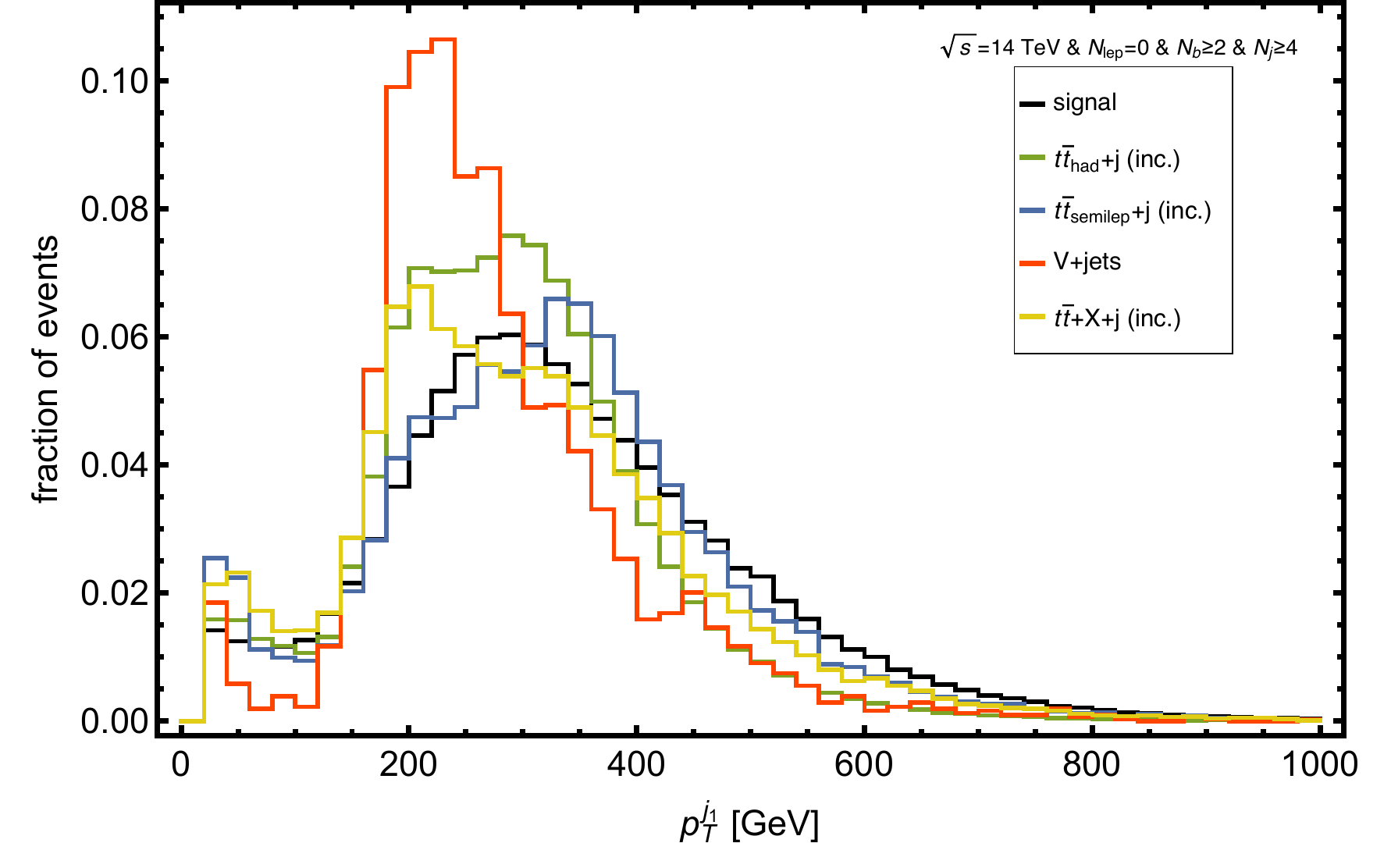} \\
			\includegraphics[scale=0.4]{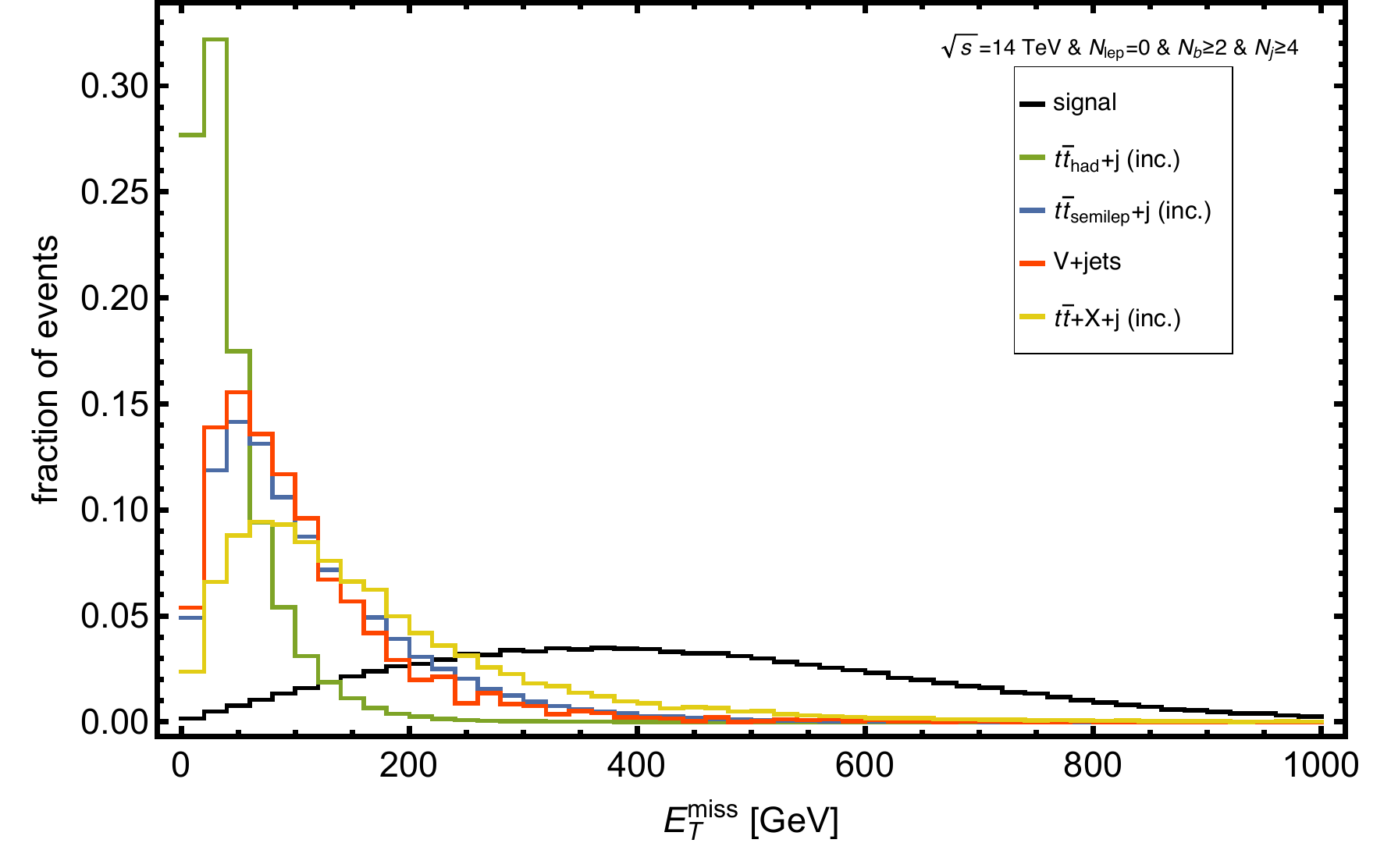} &
			\includegraphics[scale=0.4]{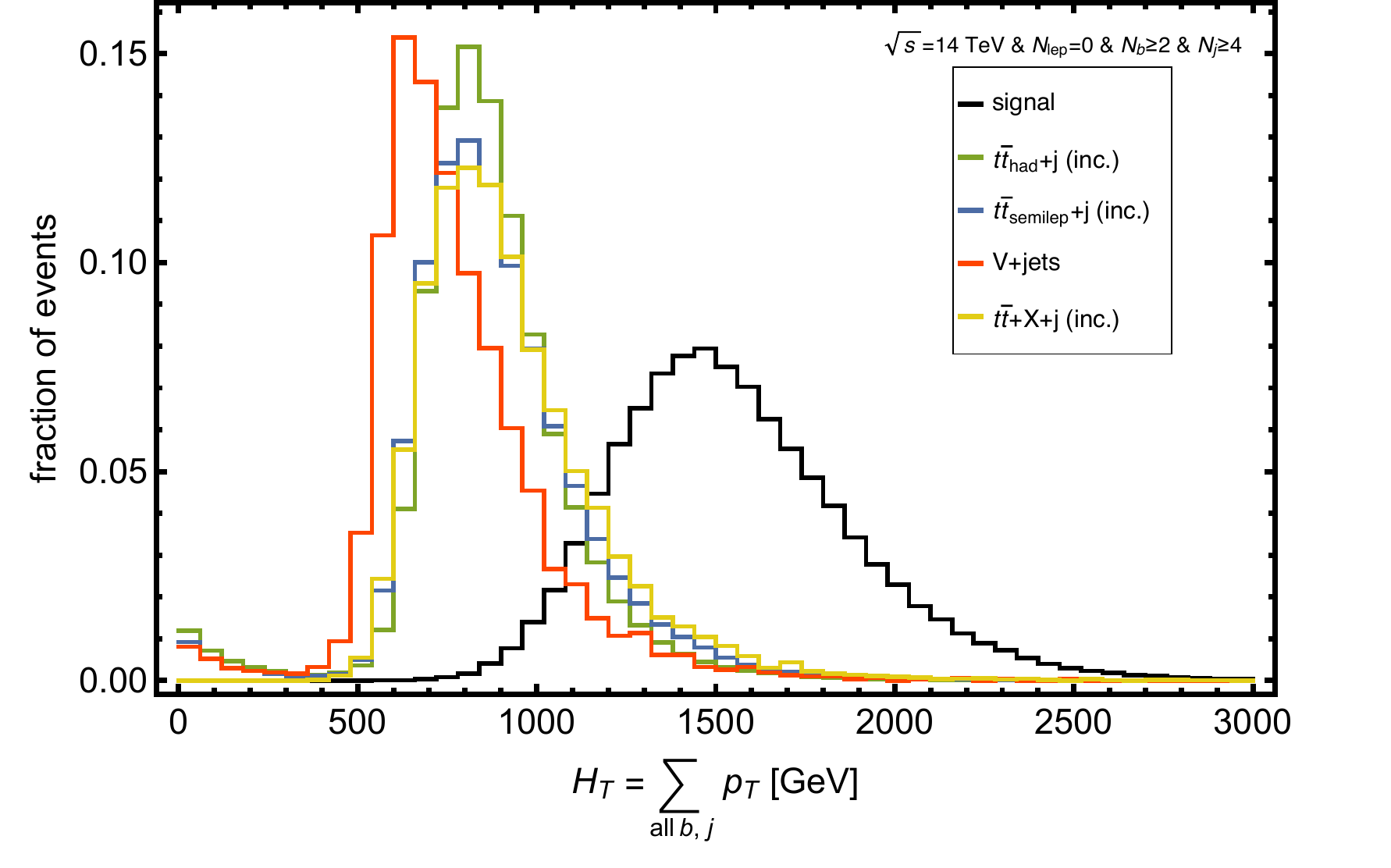} \\
			\includegraphics[scale=0.4]{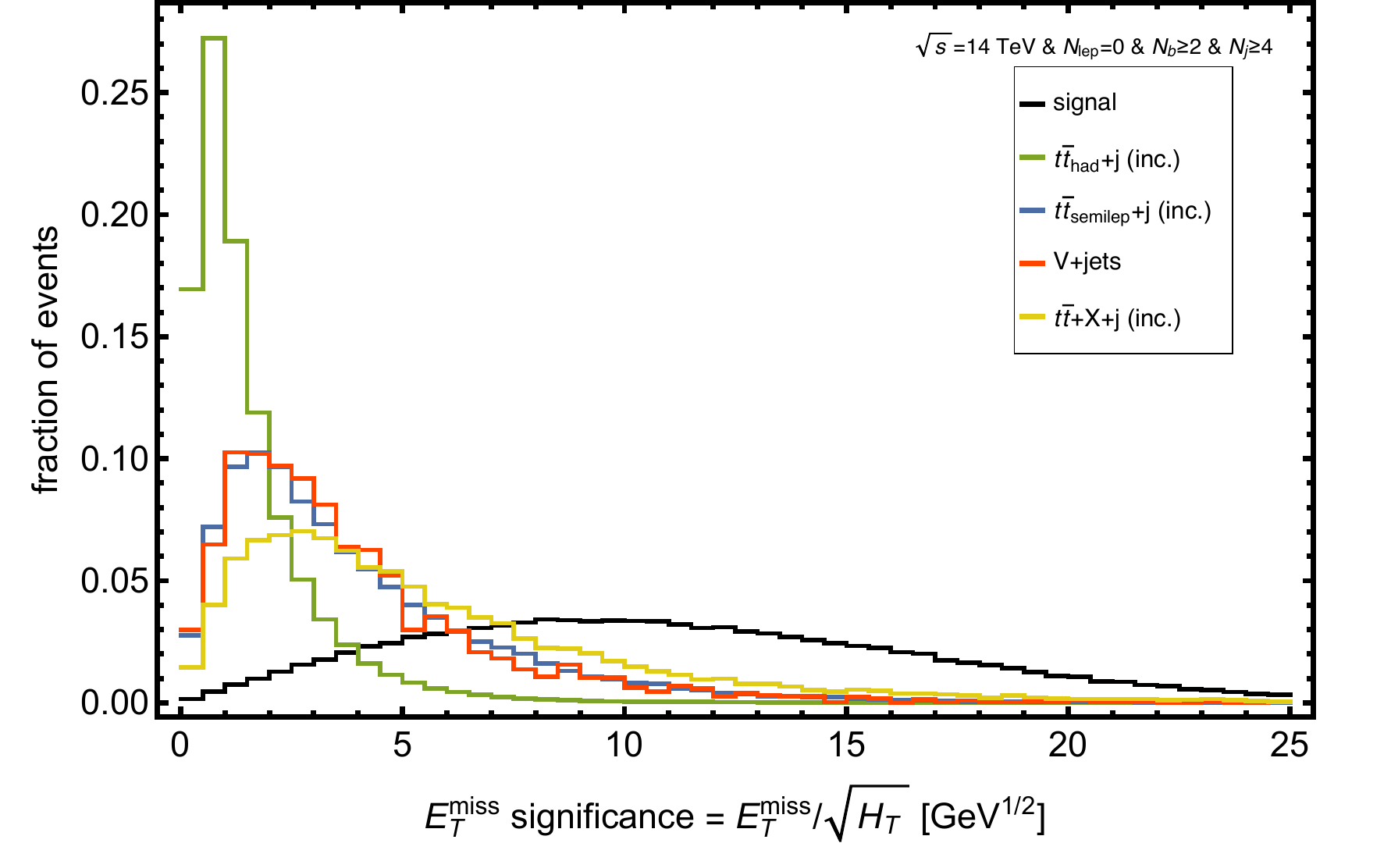} &
			\includegraphics[scale=0.4]{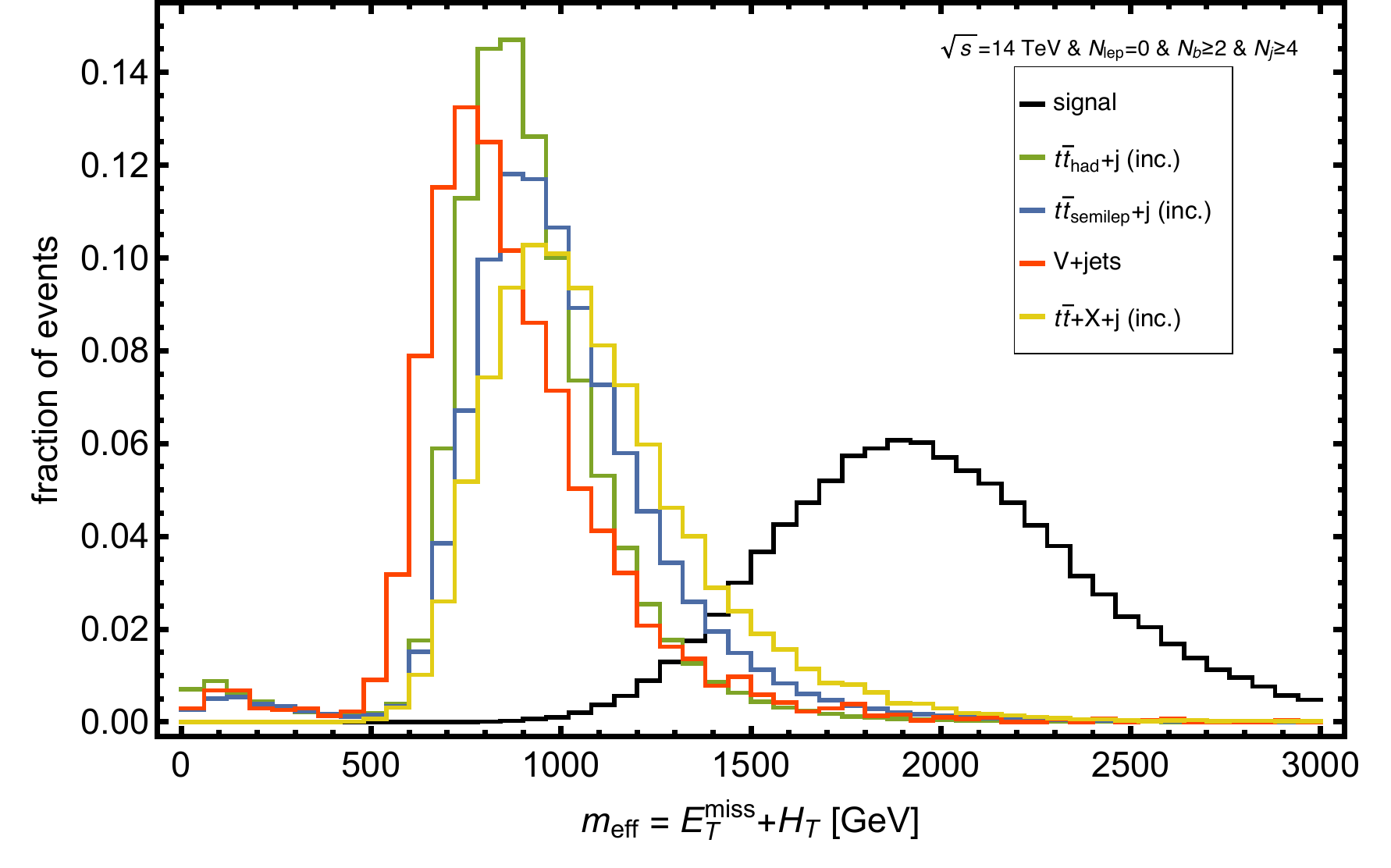}
		\end{tabular}
		\caption{\it Distributions (with a lepton veto, at least 2 $b$-jets and 4 light jets) of the fraction of signal and background events of the transverse momentum of leading $b$-jet $p_T^{b_1}$ (upper left panel), the transverse momentum of the leading light jet $p_T^{j_1}$ (upper right panel), the missing transverse energy $E_T^\text{miss}$ (medium left panel), the hadronic activity $H_T$ (medium right panel), the $E_T^\text{miss}$ significance (lower left panel), and the effective mass $m_\text{eff}$ (lower right panel).}
		\label{fig:pT-MET}
	\end{center}
\end{figure}

Fig.~\ref{fig:pT-MET} is devoted to the distributions of the fraction of signal and background events of six crucial kinematic variables: the transverse momentum of the leading $b$-jet $p_T^{b_1}$ (upper left panel); the transverse momentum of the leading light jet $p_T^{j_1}$ (upper right panel); the missing transverse energy $E_T^\text{miss}$ (medium left panel); the hadronic activity $H_T$ (medium right panel), defined as the scalar sum of the transverse momentum of all the jets ($H_T$ = $\sum_{\text{all} \, b, j}$ $p_T$); the $E_T^\text{miss}$ significance (lower left panel), which is the ratio of the missing transverse energy over the hadronic activity ($E_T^\text{miss}/\sqrt{H_T}$); and the effective mass $m_\text{eff}$ (lower right panel), defined as the sum of the missing transverse energy plus the hadronic activity ($m_\text{eff}$ = $E_T^\text{miss} + H_T$). We clearly see that the $p_T^{b_1}$  distributions for the background events have their maximum around 100 GeV, with a sharp drop after that. It is also easy to check that the $p_T^{b_1}$ distribution for the signal is less choppy, with its maximum around 500 GeV. Recall also here that the simulation of the backgrounds has been performed with the generator-level cuts, while the signal events have been simulated with only the default cuts. Therefore, a severe cut on $p_T^{b_1}$ will help to greatly reduce the background events, without affecting the signal events too much. On the other hand, a priori no similar conclusion can be drawn about the $p_T^{j_1}$ distributions of the backgrounds, which mimic the signal distribution very well. However, we will see later when we define our search strategy, that the cuts on the $p_T$ of the four leading light jets remove a large number of background events. The $E_T^\text{miss}$ distribution for the signal is practically flat (in the range from 200 GeV to 600 GeV, more or less), while for the backgrounds it peaks below 100 GeV and drops sharply thereafter, with very little fraction of events above 200 GeV. It is therefore to be expected that a cut around this value eliminates much of the background events without much change in the number of signal events. In addition, our signal presents a significant peak around 1500 GeV for the hadronic activity distribution, while the peaks of the $H_T$ distributions for the backgrounds are below 1000 GeV, with very little fraction of events above this value. Again, an $H_T$ cut at 1000 GeV and above should be very useful for getting the backgrounds out of the way and keeping a large proportion of signal events. $E_T^\text{miss}$ significance distributions for the backgrounds are mostly below 5, with peaks around values of 2-3. The signal distribution, however, is much less steep, being more or less flat between 5 and 15. From this we can also conclude that a $E_T^\text{miss}$ significance cut above 5 should be very helpful in reducing the backgrounds without affecting the signal. Finally, the effective mass $m_\text{eff}$ also appears to be a very efficient variable for separating signal from background. The signal distribution peaks around 1800 GeV while the background ones have peaks around 700-800 GeV, with very few events beyond 1300 GeV.

All these six kinematic variables, shown in Fig.~\ref{fig:pT-MET}, together with the transverse momenta of the subleading light jets and $b$-jets, not shown here for space saving, indicate in general a very distinct behavior between signal and background. This motivates the definition of our search strategy, through the cuts shown below, separating into two signal regions: a first signal region (SR1) in which we ask for at least two $b$-jets in the final state and another one (SR2) with at least three reconstructed $b$-jets.
Also, both signal regions require at least four light jets. The $p_T$ cuts at detector level for all  the jets are then:
\begin{eqnarray}
&&p_T^{j_1} > 200 \, \text{GeV} \,, \quad p_T^{j_2} > 150 \, \text{GeV} \,, \quad p_T^{j_3} > 80 \, \text{GeV} \,, \quad p_T^{j_4} > 40 \, \text{GeV} \,, \nonumber\\
&&loose:\quad p_T^{b_1} > 100 \, \text{GeV} \,, \quad p_T^{b_2} > 60 \, \text{GeV} \,, \nonumber\\
&&tight:\quad p_T^{b_1} > 100 \, \text{GeV} \,, \quad p_T^{b_2} > 60 \, \text{GeV} \,, \quad p_T^{b_3} > 35 \, \text{GeV} \,.
\label{pTcuts}
\end{eqnarray}
Based on the above, we define the SR1 search strategy with the following cuts:
\begin{itemize}
\item Loose selection cuts of Eq.~(\ref{SRdefinitions}),
\item loose $p_T$ cuts of Eq.~(\ref{pTcuts}),
\item $E_T^\text{miss}>$ 150 GeV,
\item and $m_\text{eff}>$ 1800 GeV,
\end{itemize}
whilst the SR2 search strategy has these cuts:
\begin{itemize}
\item Tight selection cuts of Eq.~(\ref{SRdefinitions}),
\item tight $p_T$ cuts of Eq.~(\ref{pTcuts}),
\item $E_T^\text{miss}>$ 150 GeV,
\item and $m_\text{eff}>$ 1300 GeV.
\end{itemize}
In order to study the potential of our search strategies, we are going to make use of the following expression for the statistical significance of the number of signal events, $S$, with respect to the number of background events, $B$~\cite{Cowan:2010js,Cowan:2012}:
\begin{equation}
\mathcal{S}_\text{sta}=\sqrt{-2 \left((B+S) \log \left(\frac{B}{B+S}\right)+S\right)} \,.
\label{statS}
\end{equation}
In addition, to obtain a more realistic estimate of the significances~\footnote{Using the {\tt Zstats} package~\cite{Zstats}, we have verified that the significances obtained with Eqs.~(\ref{statS}) and~(\ref{systS}) are compatible with the values obtained with the expressions for discovery significances proposed in~\cite{Kumar:2015tna,Bhattiprolu:2020mwi}, with differences of at most 5\%.}, we can take background systematic uncertainties into account by modifying Eq.~(\ref{statS}) as follows~\cite{Cowan:2010js,Cowan:2012}:
\begin{equation}
{\cal S}_\text{sys} = \sqrt{2 \left((B+S) \log \left(\frac{(S+B)(B+\sigma_{B}^{2})}{B^{2}+(S+B)\sigma_{B}^{2}}\right)-\frac{B^{2}}{\sigma_{B}^{2}}\log \left(1+\frac{\sigma_{B}^{2}S}{B(B+\sigma_{B}^{2})} \right) \right)} \,,
\label{systS}
\end{equation}
where $\sigma_{B}=(\Delta B) B$, with $\Delta B$ being the relative systematic uncertainty, that we choose to be, in a conservative way, of 30\%.

\begin{table}
\hspace*{-12.5mm}
    \centering
\begin{tabular}{r|rrrrr|cc}
\hline\hline
   Process  & signal & $t\bar{t}_{\rm had}+j$ (inc.) & $t\bar{t}_{\rm semilep}+j$ (inc.) & $V$+jets & $t\bar{t}X+j$ (inc.) & $\cal{S}_\text{sta}$ & $\cal{S}_\text{sys}$ \\
   \hline
    Expected  & 20 & $2.19 \times 10^6$ & $0.67 \times 10^6$ & $3.56 \times 10^5$ & $2.9\times 10^3$ & $0.01$ & $2\times 10^{-5}$ \\
    \hline
    selection cut  & 15.7 & $2.98\times 10^5$ & $2.6\times 10^4$ & 4435 & 505.5 & 0.03 & $1.5\times 10^{-4}$ \\
    loose $p_T$ cuts  & 7.7 & 7341 & 259.3 & 12.7 & 14.3 & 0.09 & $3.3\times 10^{-3}$ \\
    $E_T^\text{miss}>150$ GeV  & 7.1 & 60.9 & 37.8 & 0 & 5.1 & 0.68 & 0.21 \\
    $m_\text{eff}>1800$ GeV & 5.5 & 1.0 & 1.5 & 0 & 0.2 & 2.69 & 2.30 \\
    \hline\hline
\end{tabular}
    \caption{\it Cut flow for SR1. Loose selection cuts shown in Eq.~\eqref{SRdefinitions} and $p_T$ cuts of Eq.\eqref{pTcuts}. Significances from Eqs.~(\ref{statS}) and~(\ref{systS}), the latter with a background systematic uncertainty of 30\%.}
    \label{cutflowSR1}
\end{table}

\begin{table}
\hspace*{-12.5mm}
    \centering
\begin{tabular}{r|rrrrr|cc}
\hline\hline
   Process  & signal & $t\bar{t}_{\rm had}+j$ (inc.) & $t\bar{t}_{\rm semilep}+j$ (inc.) & $V$+jets & $t\bar{t}X+j$ (inc.) & $\cal{S}_\text{sta}$ & $\cal{S}_\text{sys}$ \\
   \hline
    Expected  & 20 & $2.19 \times 10^6$ & $0.67 \times 10^6$ & $3.56 \times 10^5$ & $2.9\times 10^3$ & $0.01$ & $2\times 10^{-5}$ \\
    \hline
    selection cut  & 9.8 & $2.78 \times 10^4$ & 1841 & 145.7 & 94.1 & 0.06 & $1.1\times 10^{-3}$ \\
    tight $p_T$ cuts  & 4.4 & 197.1 & 3.7 & 0 & 2.1 & 0.31 & 0.07 \\
    $E_T^\text{miss}>150$ GeV  & 4 & 1.9 & 0.7 & 0 & 0.4 & 1.95 & 1.66 \\
    $m_\text{eff}>1300$ GeV & 3.9 & 0 & 0.4 & 0 & 0 & 3.51 & 3.34 \\
    \hline\hline
\end{tabular}
    \caption{\it Cut flow for SR2. Tight selection cuts shown in Eq.~\eqref{SRdefinitions} and $p_T$ cuts of Eq.\eqref{pTcuts}. Significances from Eqs.~(\ref{statS}) and~(\ref{systS}), the latter with a background systematic uncertainty of 30\%.}
    \label{cutflowSR2}
\end{table}

We are now in a position to apply our search strategies on the events of our signal and the backgrounds generated for an LHC energy of 14 TeV and a total integrated luminosity of 1000 fb$^{-1}$. In Tabs.~\ref{cutflowSR1} and~\ref{cutflowSR2} the cut flow of the the SR1 and SR2 signal regions are shown, respectively, together with their corresponding significances as we apply each of the cuts. In the SR1 case (Tab.~\ref{cutflowSR1}), we see that the selection cuts reduce more than one order the magnitude all the background events, while keeping the 75\% of the signal events. In this signal region, the loose $p_T$ cuts are very efficient, reducing backgrounds by more than two orders of magnitude and only half the signal. The $E_T^\text{miss}$ cut is also very useful, eliminating most of the $t \bar t$ and $t \bar t + X$ events and bringing the $V$+jets background to zero, while barely affecting the signal events. Finally, the $m_\text{eff}$ variable eliminates most of the $t \bar t$-like events, leaving only 2.7 events of the total $t \bar t$ background and keeping 5.5 signal events, more than 25\% of those initially expected. This all adds up to a final statistical significance close to the evidence level and somewhat greater than 2 when considering 30\% systematic uncertainties in the background. The results for the SR2 search strategy are more stimulating, as shown in the cut flow of Tab.~\ref{cutflowSR2}. The tight selection cuts reduce the hadronic $t \bar t$ background by two orders of magnitude and all other backgrounds by more than three orders of magnitude, while keeping half of the signal events. The $p_T$ cuts eliminate the $V$+jets background and again reduce the remaining backgrounds by more than two orders of magnitude, with half of the remaining signal events surviving. The $E_T^\text{miss}$ cut again hardly affects the signal, reduces by two orders of magnitude the events of the hadronic $t \bar t$ background, which are finally removed by the $m_\text{eff}$ cut, which hardly modifies the signal, eliminates the $t \bar t + X$ events and leaves the only surviving background in this signal region, semileptonic $t \bar t$, at 0.36 events. In the end, in this signal region we obtain for both significance estimates values above the evidence level. At this point, it is important to note that in both signal regions the cuts can be further adjusted, preserving at least three signal events and killing all the simulated backgrounds at the same time. For instance, for SR1 (SR2) with $m_\text{eff} >$ 2100 GeV ($m_\text{eff} >$ 1500 GeV), 3.2 (3.7) signal events remain and the background events vanish. Notice that this kinematic variable summarizes the main feature of our signal, with several energetic light jets and $b$-jets, that differs from the more conventional ones (with full decays to the LSP).

The projections for a luminosity of 3000 fb$^{-1}$, considering that the number of signal and background events increase in the same way, are very promising. For the SR1 search strategy we obtain $\cal{S}_\text{sta}$ = 4.66 and $\cal{S}_\text{sys}$ = 3.23, and $\cal{S}_\text{sta}$ = 6.08 and $\cal{S}_\text{sys}$ = 5.32 for the SR2 case. That is, for the future high-luminosity phase of the LHC, one could expect significances above the evidence level in the SR1 signal region and reach significances larger than the discovery level with the SR2 search strategy, which shows that this class of experimental signatures at the LHC deserve special attention and dedicated searches.

%%%%%%%%%%%%%%%%%%%%%%%%%%%%%%%%%%%%%%%%%%%%%%%%%%%%%%%%%%%%%%%%

\section{Conclusions}
\label{conclus}

In this work we have developed a proof-of-concept collider analysis at the HL-LHC for a new SUSY signal (whose spectrum evades current LHC searches): $pp \to \tilde g \tilde g$ $\to (\tilde \chi_3^0 j j) \, (\tilde \chi_3^0 j j)$ $\to (\tilde \chi_1^0 h j j) \, (\tilde \chi_1^0 h j j)$ $\to 4j$ + $4 b$ + $E_T^\text{miss}$. The more problematic SM backgrounds of this experimental signature are $t \bar t$, $V$+jets ($V$= $W$, $Z$), and $t \bar t + X$ ($X$ = $W$, $Z$, $\gamma^*$, $h$), which all turn out to be under control after the cuts of our search strategy. The selection cuts define two signal regions, SR1 with $N_b \geq$ 2 and SR2 with $N_b \geq$ 3, to which we subsequently applied cuts on the most relevant kinematic variables: the transverse momenta of light and $b$-jets, $E_T^\text{miss}$, and $m_\text{eff}$, which is the sum of $E_T^\text{miss}$ plus the hadronic activity, $H_T$. With a center-of-mass energy of 14 TeV and a total integrated luminosity of 1000 fb$^{-1}$ we reach signal significances close to the evidence level (3$\sigma$) for SR1 and above this value for SR2. The prospects for 3000 fb$^{-1}$ are very encouraging, with significances greater than 3$\sigma$ for SR1 and above the discovery level (5$\sigma$) for SR2, indicating that this novel signature deserves the development of dedicated searches by the LHC experiments.

\section*{Acknowledgments}
The work of EA and RM is partially supported by the ``Atracci\'on de Talento'' program (Modalidad 1) of the Comunidad de Madrid (Spain) under the grant number 2019-T1/TIC-14019 and by the Spanish Research Agency (Agencia Estatal de Investigaci\'on) through the grant IFT Centro de Excelencia Severo Ochoa SEV-2016-0597 (EA, RM). The work of EA is also partially supported by CONICET and ANPCyT under projects PICT 2016-0164, PICT 2017-2751, and PICT 2017-2765. The work of AD was partially supported by the National Science Foundation under grant PHY-1820860. The work of MQ is partly supported by Spanish MINEICO under Grant FPA2017-88915-P, by the Catalan Government under Grant 2017SGR1069, and by Severo Ochoa Excellence Program of MINEICO under Grant SEV-2016-0588. IFAE is partially funded by the CERCA program of the Generalitat de Catalunya.

\bibliographystyle{JHEP}
\bibliography{lit}

\end{document}